\documentclass[12pt,aps,showpacs]{revtex4}
\usepackage{graphicx}
\begin{document}
\title{Quantum logic operations and creation of entanglement
in a scalable superconducting quantum computer with long-range
constant interaction between qubits}
\author{G. P. Berman$^1$, A. R. Bishop$^1$, D. I. Kamenev$^{1}$,
and A. Trombettoni$^{1,2}$}
\affiliation{$^1$Theoretical Division,
Los Alamos National Laboratory, Los Alamos,
New Mexico 87545}
\affiliation{$^2$ I.N.F.M. and Dipartimento di Fisica, Universit\`a
di Parma, parco Area delle Scienze 7A
Parma, I-43100, Italy}

\begin{abstract}
We consider a one-dimensional chain of many superconducting
quantum interference devices (SQUIDs), serving as charge qubits.
Each SQUID is coupled to its nearest neighbors through  constant
capacitances. We study the quantum logic operations and
implementation of entanglement in this system.
 Arrays with two and three qubits are considered in
detail.  We show that the creation of entanglement with an
arbitrary number of qubits can be implemented, without systematic
errors, even when the coupling between qubits is not small. A
relatively large coupling constant allows one to increase the
clock speed of the quantum computer. We analytically and
numerically demonstrate the creation of the entanglement for this
case, which can be a good test for the experimental implementation
of a relatively simple quantum protocol with many qubits. We
discuss a possible application of our approach for implementing
universal quantum logic for more complex algorithms by decreasing
the coupling constant and, correspondingly, decreasing the clock
speed. The errors introduced by the long-range interaction for the
universal logic gates are estimated analytically and calculated
numerically. Our results can be useful for experimental
implementation of quantum algorithms using controlled magnetic
fluxes and gate voltages applied to the SQUIDs. The algorithms
discussed in this paper can be implemented using already existing
technologies in superconducting systems with constant inter-qubit
coupling.
\end{abstract}



\maketitle

\section{Introduction}

In the last few years, a large number of studies has been devoted
to the realization of qubits using Josephson
devices~\cite{makhlin01}. A single superconducting qubit can be
realized using either charge or flux degrees of freedom. The flux
Josephson two-state system is based on two quantum states carrying
opposite persistent currents. Coherent time evolution between
these states has been recently observed~\cite{chiorescu03}. At
present, however, no quantum oscillations between two coupled flux
qubits has been reported. In charge Josephson qubits the relevant
degree of freedom is the charge on superconducting grains. The
coherent control of macroscopic quantum states in a single
Cooper-pair box has been demonstrated~\cite{nakamura99}, while
the first observation of coherent quantum oscillations in two
coupled charge qubits has been reported~\cite{pashkin03}.
Moreover, coherent oscillations have been observed in other
superconducting devices~\cite{ramos,martinis02,yu02}. While longer
coherence times are desirable, these experiments show that the
superconducting circuits are strong candidates for solid-state
qubits. The next major step toward building a Josephson-junction
based quantum computer is to experimentally realize simple quantum
algorithms, such as the creation of an entangled state involving
more than two coupled qubits.

A typical design of a Cooper pair box consists of a small
superconducting island with $n$ Cooper pair charges connected by a
tunnel junction with a Josephson coupling energy $E_J$ and the
capacitance ${\rm C}_J$ to a superconducting
electrode~\cite{makhlin01}. A control gate voltage $V$ is
coupled to the system via a gate capacitor ${\rm C}_g$.

The energy of the $n$ Cooper pairs in the box is $E=4 E_{\rm
C}n^2-2enV$, where $E_{\rm C}=e^2/2({\rm C}_g+{\rm C}_J)$, and
$e$ is the electron charge. Considering later $n$ as one of the
canonical variables, and neglecting the constant term, we can
represent $E$ in the form $E=4 E_{\rm C}(n-n_g)^2$, where
$n_g={\rm C}_g V/2e$ is the gate charge (in units of $2e$). When
$E_{\rm C} \gg E_J$, by choosing $n_g$ close to the degeneracy
point, $n_g=(2n+1)/2$, only the states with $n$ and $n+1$ Cooper
pairs play a role. In this case, the effective Hamiltonian of the
two-state system can be written in the spin-$\frac 12$ notation as
\begin{equation}
\label{one_spin_Hamiltonian} \bar\mathbf{H}_1=-\bar B^z
\mathbf{S}^z-B^x\mathbf{S}^x,
\end{equation}
where the state with $n$ Cooper pairs corresponds to the spin
state $\left(\matrix{1 \cr 0 \cr}\right)$ and the state with
$n+1$ Cooper pairs corresponds to the spin state
$\left(\matrix{0 \cr 1 \cr}\right)$
(see, for example,~\cite{makhlin01});
$\mathbf{S}^z$ and $\mathbf{S}^x$ are, respectively, the $z$ and
$x$ - components of spin-$\frac 12$ operator;
$\bar B_z \sim 4E_c(1-2n_g)$
and $B_x \sim E_J$ are the effective magnetic fields which are
controlled by the applied gate voltage and the magnetic flux. [See
Eqs. (\ref{BzBx}) below.]

The Hamiltonian of an array of coupled superconducting
qubits can be written in the general form
\begin{equation}
\label{many_spins_Hamiltonian} \bar\mathbf{H} = -
\sum_{i=0}^{N-1} \bar B^z_i \mathbf{S}^z_i - \sum_{i=0}^{N-1} B^x_i
\mathbf{S}^x_i + \sum_{i,j=0\atop i \ne j}^{N-1}
U_{ij}^{a} \mathbf{S}^a_i \mathbf{S}^b_j,
\end{equation}
where $a=x,y,z$ and $N$ is the number of qubits in the circuit.
The explicit form of the coefficients $U^{a}_{ij}$ in
(\ref{many_spins_Hamiltonian}) depends on the particular way in
which the inter-qubit coupling is implemented. A variety of
possible architectures to couple charge Josephson qubits has been
proposed~\cite{makhlin99,plastina01,you01,you02,
blais03,johnson03,plastina03}. It has been
suggested~\cite{makhlin99} to realize a coupling of the type $\propto
\mathbf{S}^y_i \mathbf{S}^y_j$ using an inductor: all qubits were
connected in parallel to an LC-oscillator which provided the
two-qubit interaction. A possible limitation of this architecture
is that the inter-qubit coupling term is valid only under the
conditions that the phase conjugate to the total charge on the
qubit capacitors fluctuates weakly, and that the eigenfrequency
$\omega_{0}$ of the LC circuit is much larger than the typical
frequencies of the qubit dynamics. Since $\omega_0$ scales with
the number $N$ of qubits as $1/\sqrt{N}$, this limits the maximum
number of qubits in the quantum register. A different proposed
type of the inter-qubit coupling uses the Coulomb interaction
between charges on the islands of the charge
qubits~\cite{plastina01}: this gives a coupling $\propto
\mathbf{S}^z_i \mathbf{S}^z_j$. A drawback of this approach is
that one cannot switch the inter-qubit coupling in this scheme
without introducing unwanted dephasing effects. Another proposed
architecture employs two additional SQUIDs to connect each qubit:
all Josephson charge qubits are coupled through a common
superconducting inductance, and the resulting coupling is $\propto
\mathbf{S}^x_i  \mathbf{S}^x_j$~\cite{you02}. The same coupling of
charge qubits can also be realized using mutual
inductance~\cite{you01}. Finally, we mention that another
experimental realization of a single superconducting charge qubit
has been obtained using the current-biased Josephson 
junctions~\cite{martinis02,yu02}. A quantum computer 
architecture based on two capacitively coupled
current-biased Josephson junctions was discussed 
before~\cite{blais03,johnson03,plastina03}.

In this paper, we consider capacitively coupled SQUIDs, which
currently are the only experimental setup that has provided an
experimental detection of quantum oscillations in two coupled
charge qubits~\cite{pashkin03}. In this architecture, the SQUIDs
are connected via a constant capacitor, and the coupling is of the
form $\propto \mathbf{S}^z_i \mathbf{S}^z_j$. 
Varying the magnetic flux $\Phi$ through the
SQUIDs allows one to control the Josephson energy $E_J$, which is
equivalent to controlling the effective magnetic field
$B^x$~\cite{makhlin01}. Therefore, we will assume that it is
possible to have the Josephson energies [{\it i.e.}, $B^x_i$
coefficients in Eq.(\ref{many_spins_Hamiltonian})] independently
and locally variable in time. 

We note that experimentally the 
fast independent manipulation of gate voltages is 
possible~\cite{pashkin1,pashkin2}. At the same time, the fast 
independent control of magnetic fluxes through the different 
SQUIDs with the distance $\sim 1~\mu$m between them and the 
switching time $\sim 1$~ns 
represents a challenge to the present-day technology.
As an alternative, the $B^x$ field can be manipulated 
by using recently proposed~\cite{stacked} Nb/AlO$_x$Nb-Al/AlO$_x$/Nb
stacked Josephson junctions instead of SQUIDs. 
In the double Josephson junction device, 
injection current $I_{inj}$ (of the order of several mA)
in the control junction induces 
variation of the Josephson critical current $I_c$
(of the order of several $\mu$A). As
for a single Josephson junction $B_x\sim E_J=I_c\Phi^0/(2\pi c)$
($\Phi^0=hc/2e$ is the flux quantum,
$h$ is the Planck's constant, and $c$ is the light velocity)
manipulation of $I_c$ allows one to control the $B^x$ field. 
In the experiment~\cite{stacked} $I_c=0$ (and $B^x=0$) 
for $I_{inj}\approx 2.5$ mA. Using the electrical signals instead 
of the localized magnetic fluxes for conrolling 
the $B^x$ field would allow one to speed up the quantum computer 
operations and to simplify the design. The results presented 
in this paper are applicable also to the quantum computer based 
on the stacked Josephson junctions (instead of SQUIDs)
serving as qubits.  
 
\begin{figure}
\centerline{\includegraphics[width=7cm,height=9cm,angle=270]{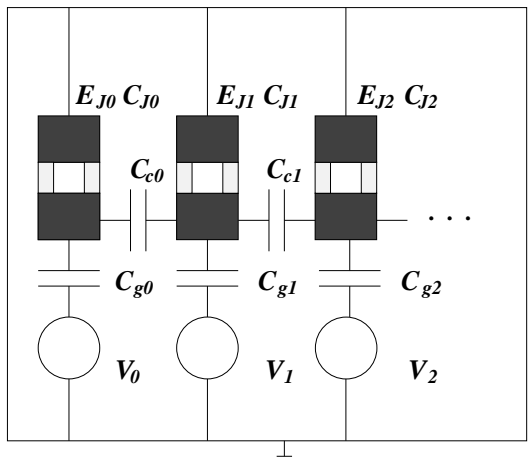}}
\vspace{-5mm}
\caption{A schematic illustration of
array of capacitively coupled SQUIDs.}
\label{fig:1}
\end{figure}

\section{Array of capacitively coupled SQUIDs
 as a quantum register}

A schematic plot of an array of $N$ capacitively coupled SQUIDs is
shown in Fig.~1. The $i$th SQUID with $i=0,\cdots,N-1$ corresponds
to the $i$th qubit. The SQUIDs have Josephson energies $E_{Ji}$
and capacitances ${\rm C}_{Ji}$. Each SQUID is connected to the
control gate voltages $V_i$ via a gate capacitor ${\rm C}_{gi}$.
The $i$th intermediate qubit is connected to its two neighboring
($i \pm 1$)th qubits via the capacitors ${\rm C}_{i,i \pm 1}$,
where ${\rm C}_{i,i+1}$ are the off-diagonal elements of the
capacitance matrix. In Fig.~1 we use the notation ${\rm C}_{ci} =
{\rm C}_{i,i+1}$. The end $0$th and $(N-1)$th SQUIDs are connected
to one, respectively, the $1$st and $(N-2)$th SQUIDs. The
Hamiltonian corresponding to the charging energy of this system
can be written as
\begin{equation}
\mathbf{H}_{\rm C}(t)= 4
\sum_{i,j=0}^{N-1}[\mathbf{n}_i-n_{gi}(t)]\Gamma_{ij}
[\mathbf{n}_j-n_{gj}(t)],
\label{charging_energy_Hamiltonian}
\end{equation}
where $\mathbf{n}_i$ is the operator for the total number of
Cooper pairs in the $i$th SQUID, and
$n_{gi}(t)={\rm C}_{gi} V_i(t)/2e$
is the charge (in units of $2e$) induced on the $i$-th qubit
by the corresponding dc electrode;
$\Gamma_{ij}$ is related to the inverse of the
capacitance matrix ${\rm C}_{ij}$ of the system by
\begin{equation}
\Gamma_{ij}=\frac{e^2}{2} {\rm C}_{ij}^{-1}. \label{u}
\end{equation}
To illustrate the structure of the capacitance matrix we assume
that all SQUIDs are identical with the same capacitances
${\rm C}_{gi}={\rm C}_{g}$, ${\rm C}_{Ji}={\rm C}_{J}$, and ${\rm
C}_{ci}={\rm C}_c$. (This assumption is made for this example only.)
Then, the capacitance matrix is
\begin{equation}
{\rm C}_{ij}={\rm C}_0 \left[\delta_{ij}(1+g\eta)-
\eta\sum_d\delta_{i,i+d}\right], \label{c}
\end{equation}
where ${\rm C}_0={\rm C}_g+{\rm C}_J$, $\eta={\rm C}_c/{\rm C}_0$;
$\delta_{ij}$ is the Kronecker delta-symbol;
$d=\pm 1$ and $g=2$ for the intermediate qubits;
$d=1$ and $g=1$ for the $0$th qubit; $d=-1$ and $g=1$ for the
last $(N-1)$th qubit. If other devices (for measurement)
are attached to the SQUIDs, their
capacitances must be added to C$_0$.

Since the matrix $\Gamma_{ij}$ is the inverse of a tridiagonal
matrix, it has nonzero matrix elements on the second, third
and other
diagonals which characterize the long-range interaction between
the qubits. However, for $\eta\ll 1$ the off-diagonal elements of
$\Gamma_{ij}$ decay exponentially as $\eta^{|i-j|}$, so that the
influence of the long-range interaction can be reduced by taking
the coupling capacitances ${\rm C}_c$ to be much smaller than the
on-site capacitances ${\rm C}_0$. (In the experiment~\cite{pashkin03} 
the value of $\eta$ was chosen to be $\eta \sim 0.05$.)

Usually, long-range interaction between qubits is difficult to
control, and this interaction represent a problem for quantum
computation. However, in the case when there are only two states
in the quantum register, we will show that the destructive effect
of the long-range interaction can be completely suppressed by the
proper choice of protocol parameters, so that the condition
$\eta\ll 1$ is not required in this case. Since the clock speed of
the quantum computer is proportional to $\eta$, increasing the
value of $\eta$ would allow one to increase the speed of
implementation of a quantum algorithm. This particular example
with two states can be important for benchmarking a scalable
superconducting quantum computer device.

In order to obtain the effective Hamiltonian in
spin-$\frac{1}{2}$ notation (see, for
example,~\cite{bruder93,glazman97}), we make the substitution
$n_{gi}=n^0_i+m_{gi}$ and $\mathbf{n}_{i}=n^0_i+\mathbf{m}_{i}$,
where $\mathbf{m}_{i}$ is the operator of excess number of
Cooper pairs on $i$th SQUID. The dimensionless charging
Hamiltonian~(\ref{charging_energy_Hamiltonian}) becomes
\begin{equation}
\mathbf{H}_{\rm C}^\prime(t)=
{{\rm C}_0\over 2e^2}{\cal \mathbf{H}}_{\rm C}(t)
=\sum_{i,j=0}^{N-1} [\mathbf{m}_i-m_{gi}(t)] U_{ij}
[\mathbf{m}_j-m_{gj}(t)], \label{HC}
\end{equation}
where $U_{ij}=\Gamma_{ij}{\rm C}_0/(2e^2)$ and ${\rm C}_0$
is the on-site capacitance for the $0$th qubit.
It is convenient to
introduce the spin operator $\mathbf{S}_{i}^z=1/2-\mathbf{m}_i$
with projections $s_i^z=\pm 1/2$ on the $z$-axis.
We treat $\mathbf{m}_i$ as an operator with eigenvalues $0$ and $1$.
Then, the state with the eigenvalue $m_i=0$ corresponds to the
spin state $\left(\matrix{1 \cr 0 \cr}\right)$, and the state with
the eigenvalue $m_i=1$ corresponds to the spin state
$\left(\matrix{0 \cr 1 \cr}\right)$~\cite{makhlin01}. Using the
relation between $\mathbf{m}_{i}$
($\mathbf{m}_{j}$) and $\mathbf{S}_{i}^z$ ($\mathbf{S}_{j}^z$),
the symmetry property $U_{ij}=U_{ji}$, and the fact that
$(\mathbf{S}^z_i)^2={1\over 4}\left(\matrix{1&0 \cr 0&1
\cr}\right)$, the Hamiltonian of the whole system can be written
in the form
\begin{equation}
\label{HCS} \mathbf{H}(t)=-\sum_{i=0}^{N-1}
\bar B^z_i(t)\mathbf{S}^z_i-\sum_{i=0}^{N-1}B^x_i(t)\mathbf{S}^x_i+
\sum_{{i,j=0 \atop i\ne
j}}^{N-1}U_{ij}\mathbf{S}^z_i\mathbf{S}^z_j.
\end{equation}
Here we omitted the term
$\sum_{ij=0}^{N-1}U_{ij}[m_{gi}(t)-1/2][m_{gj}(t)-1/2]+(1/4)
\sum_{i=0}^{N-1}U_{ii}$, which does not contain the spin operators
and does not influence the quantum equations of motion. The
controllable $\bar B^z$ and $B^x$ fields are expressed through
the parameters of the model:
\begin{equation}
\label{BzBx}  \bar B^z_i(t)=\sum_{j=0}^{N-1}U_{ij}[1-2m_{gj}(t)],
\qquad
B^x_i(t)={{\rm C}_0\over 2e^2}E_{Ji}[\Phi_i(t)]= {{\rm C}_0\over
e^2}E_{Ji}^0 \cos{[\pi \Phi_i(t) / \Phi^0]},
\end{equation}
where $E_{Ji}^0$ is the Josephson energy of each of the two
Josephson junctions of the $i$-th SQUID, $\Phi_i(t)$ is the magnetic flux
through the $i$-th SQUID.
From the second expression in Eq. (\ref{BzBx}) it follows that
$B^x_i=0$ for non-zero flux $\Phi_i=(2k-1)\Phi^0/2$,
where $k=1,2,\dots$.

\subsection{Two qubits}

Some basic properties of the system can be understood from an exact
analysis of the SQUID chains containing two or three qubits.
For the system of two capacitively
coupled qubits, the capacitance matrix is
\begin{equation}
{\rm C}_{ij}=\left(\matrix{{\rm C}_0+{\rm C}_c & -{\rm C}_c \cr
-{\rm C}_c & {\rm C}_1+{\rm C}_c \cr}\right)=
{\rm C}_0\left(\matrix{1+\eta & -\eta \cr
-\eta & a+\eta \cr}\right),
\label{c_2_qubits}
\end{equation}
where $a={\rm C}_1/{\rm C}_0$;
$C_0={\rm C}_{g0}+{\rm C}_{J0}$ is the on-site capacitance
for the $0$th qubit; and $C_1={\rm C}_{g1}+{\rm C}_{J1}$
is the on-site capacitance for the $1$st qubit.
The matrix $U_{ij}$ has the form
\begin{equation}
\label{J2}
U_{ij}={1\over a+(1+a)\eta}\left(\matrix{a+\eta & \eta \cr
\eta & 1+\eta \cr}\right).
\end{equation}
Since the coupling constant $U_{01}=\eta/[a+(1+a)\eta]$ is
positive, the coupling in the system is antiferromagnetic. This is
also true for larger qubit arrays.

\subsection{Three qubits}

Assuming that all on-site capacitances and coupling capacitances
are the same for all SQUIDs (we make this assumption only
for this particular example),
the capacitance matrix for the system of three qubits is
\begin{equation}
{\rm C}_{ij}=
\left(\matrix{{\rm C}_0+{\rm C}_c & -{\rm C}_c & 0 \cr
-{\rm C}_c & {\rm C}_0+2{\rm C}_c & -{\rm C}_c \cr
0& -{\rm C}_c & {\rm C}_0+{\rm C}_c \cr}\right)=
{\rm C}_0\left(\matrix{1+\eta & -\eta &0 \cr
-\eta & 1+2\eta & -\eta \cr
0 & -\eta & 1+\eta \cr}\right).
\label{c_3_qubits}
\end{equation}
The matrix $U_{ij}$ has the form
\begin{equation}
\label{J3}
U_{ij}={1\over 1+4\eta+3\eta^2}
\left(\matrix{1+3\eta+\eta^2 & \eta+\eta^2 & \eta^2 \cr
\eta+\eta^2 & 1+2\eta+\eta^2 & \eta+\eta^2 \cr
\eta^2 & \eta+\eta^2 & 1+3\eta+\eta^2 \cr
}\right).
\end{equation}
For the system of three qubits the coupling constants are
\begin{equation}
\label{J3off-diagonal} U_{01}=U_{12}={\eta+\eta^2\over
1+4\eta+3\eta^2},\qquad U_{02}={\eta^2\over 1+4\eta+3\eta^2}.
\end{equation}
As mentioned above, the off-diagonal matrix elements
$U_{i,j}$ decrease approximately as $\eta^{|i-j|}$,
where $\eta\ll 1$.
The diagonal components $U_{i,i}$, which affect
the field field $\bar B_i^z$ in the first expression in
Eq.~(\ref{BzBx}), depend on the position of the qubit in the
chain, $i$. For a system consisting of larger number of qubits, the
off-diagonal components also depend on $i$.

\section{The ${\cal \mathbf{B}}_i^z$ operator field}

It is convenient to express the Hamiltonian~(\ref{HCS})
in terms of the operator field $\mathbf{B}_i^z(t)$,
\begin{equation}
\label{Hij} \mathbf{H}(t)=-\sum_{i=0}^{N-1}
\mathbf{B}_i^z(t)\mathbf{S}^z_i
-\sum_{i=0}^{N-1}B^x_i(t)\mathbf{S}^x_i,
\end{equation}
\begin{equation}
\label{BOperator}
\mathbf{B}_i^z(t)=\sum_{j=0\atop j\ne i}^{N-1} \left\{
U_{ij}\left[1-2m_{gj}(t)-\mathbf{S}^z_j\right]\right\}+
U_{ii}[1-2m_{gi}(t)].
\end{equation}
Here different terms correspond to different sources of the
$\mathbf{B}_i^z(t)$ field in the location of $i$th qubit. The term
$U_{ij}[1-2m_{gj}(t)]$ is the field produced by the application of
the voltage to the $j$th qubit, $j\ne i$; the term
$U_{ii}[1-2m_{gi}(t)]$ is the field created by applying the
voltage directly to the $i$th qubit; and the term
$-U_{ij}\mathbf{S}^z_j$, $j\ne i$, is the field produced by the
$j$th qubit in the location of the $i$th qubit due to the constant
interaction between them.

Since the $\mathbf{B}_i^z(t)$ field is an operator, it generates
different actual $B^z_i$ fields for different states. The
$B^z_i$ field $B^z_i(p,t)$ in the location of the $i$th qubit
for the state $|p\rangle$ is defined as
\begin{equation}
\label{Bzip}
B^z_i(p,t)=\langle p|\mathbf{B}^{z}_i|p\rangle=
\sum_{j=0\atop j\ne i}^{N-1} \left\{
U_{ij}\left[1-2m_{gj}(t)-s^z_j(p)\right]\right\}+
U_{ii}[1-2m_{gi}(t)],
\end{equation}
where $s^z_j(p)$ is the eigenvalue of $\mathbf{S}^z_j$
for the state $|p\rangle$. For example, for the
state $|0\rangle=|00\dots 00\rangle$ one has $s^z_j(0)=1/2$ for
all $j$'s and
\begin{equation}
\label{Bzi0}
B^z_i(0,t)=\langle 0|\mathbf{B}^{z}_i|0\rangle=
\sum_{j=0\atop j\ne i}^{N-1} \left\{
U_{ij}\left[\frac 12-2m_{gj}(t)\right]\right\}+
U_{ii}[1-2m_{gi}(t)].
\end{equation}

Below we assume that the voltage $m_{gj}(t)$ and the control flux
$\Phi_j(t)$ ($B_j^x$ field) are varied only on one site with
$j=l$, i.e. $m_{gl}(t)=m^0_l+M_l(t)$. These parameters are
constant for all other sites, $m_{gj}(t)=m^0_j$ and
$\Phi_j(t)=\Phi^0(2k-1)/2$, $k=1,2,\dots$ for $j\ne l$.
Then the Hamiltonian
(\ref{Hij}) can be written as
\begin{equation}
\label{H01}
\mathbf{H}_l(t)=
\mathbf{H}^0+2M_{l}(t)\sum_{i=0}^{N-1}U_{li}\mathbf{S}^z_i
-B^x_l(t)\mathbf{S}^x_l,
\end{equation}
where
\begin{equation}
\label{H0}
\mathbf{H}^0=-\sum_{i=0}^{N-1}{\cal
\mathbf{B}}^{z,0}_i\mathbf{S}^z_i.
\end{equation}
Here
\begin{equation}
\label{Bz0}
\mathbf{B}^{z,0}_i=\sum_{j=0\atop
j\ne i}^{N-1} \left[
U_{ij}\left(1-2m^0_j-\mathbf{S}^z_j\right)\right]+
U_{ii}(1-2m^0_i)
\end{equation}
is the static $\mathbf{B}^z_i$ operator field. The actual static
$B^z_i$ field for a state $|p\rangle$ is
\begin{equation}
\label{Bzi0p}
B^{z,0}_i(p)=\langle p|\mathbf{B}^{z,0}_i|p\rangle=
\sum_{j=0\atop j\ne i}^{N-1} \left\{
U_{ij}\left[1-2m_{j}^0-s^z_j(p)\right]\right\}+
U_{ii}[1-2m_i^0],
\end{equation}
where the argument $p$ of $B^{z,0}_i(p)$ indicates the state
number. This field is nonuniform even for a uniform qubit chain.
To illustrate this, we consider the particular example in which all
qubits are in the same state $|0\rangle=|00\dots 00\rangle$ and
all static voltages have the same values $m^0_j=0$. Then from
Eq.~(\ref{Bzi0p}) one obtains
\begin{equation}
\label{B0im0}
B^{z,0}_i(0)=U_{ii}+\frac 12\sum_{j=0\atop j\ne
i}^{N-1}U_{ij}.
\end{equation}

\begin{figure}
\centerline{\includegraphics[width=9cm,height=9cm]{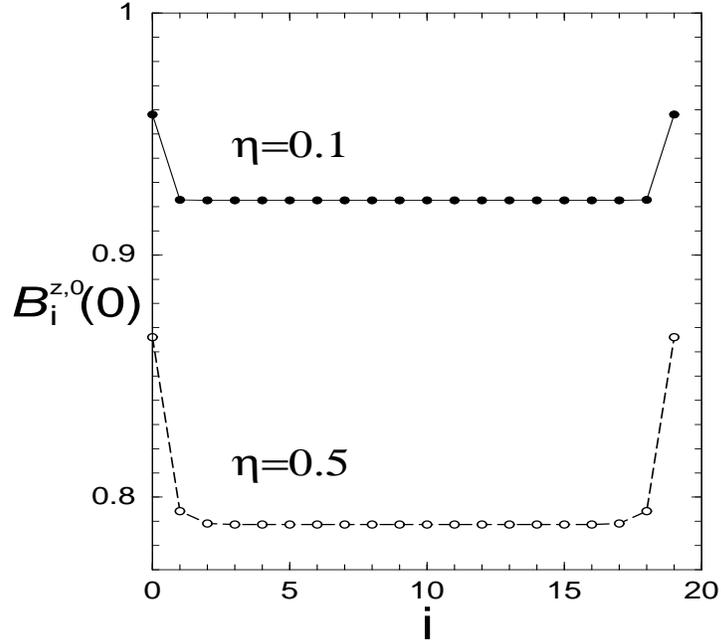}}
\vspace{-5mm} \caption{The static field $B_i^{z,0}(0)$ as a
function of the qubit number $i$ for two values of $\eta$, $m^0_j=0$
$(j=0,1,...,19)$, $N=20$.} \label{fig:2}
\end{figure}

In Fig. 2 we plot the static field $B_i^{z,0}(0)$ as a function of
the qubit number $i$ for two values of $\eta$ and for $m^0_j=0$,
$j=0,1,...,19$. The number of qubits is $N=20$. For a two-qubit
chain with the matrix (\ref{J2}) and $a=1$ one can show that
$B_0^{z,0}(0)=B_1^{z,0}(0)\approx 1-\eta/2$ for $\eta\ll 1$, which
is approximately equal to the values of $B_0^{z,0}(0)$ and
$B_{N-1}^{z,0}(0)$ in Fig. 2 for $\eta=0.1$. The static field in
Fig.~2 is nonuniform near the edges of the qubit chain despite
the fact
that all applied voltages have the same values $m^0_j=0$ and all
qubits are in the same state $|0_j\rangle$.

In dimensional
units the total static voltage applied to $i$th qubit is
$v^0_i=(2e/{\rm C}_g)(n^0_i+m^0_i)$. If $n^0_i=m^0_i=0$,
then $v^0=0$. Due to
Eq.~(\ref{HC}) the relation between the dimensional magnetic field
$B_i^{z,0}(0)$ and the effective static voltage, corresponding to
this field, can be written
as $(2e^2/{\rm C}_0)B^{z,0}_i(0)=ev_{i}^{eff}$.
For example, the value $B_i^{z,0}(0)\approx 0.92$
for $i=1,2,\dots,N-2$ and $\eta=0.1$ in Fig.~2
for ${\rm C}_0\approx 500$ aF~\cite{pashkin03} corresponds
to the effective voltage
$v_{i}^{eff}=0.92\times 2e/{\rm C}_0\approx 0.6$ mV.

\section{The interaction representation}
We decompose the wave function into the basis states $|p\rangle$
of the unperturbed Hamiltonian $\mathbf{H}^0$:
\begin{equation}
\label{psi}
\psi(t)=\sum_{r=0}^{2^N-1}c_p(t)|p\rangle
=\sum_{p=0}^{2^N-1}A_p(t)e^{-iE_pt}|p\rangle,~~~
c_p(t)=A_p(t)e^{-iE_pt},
\end{equation}
where the Planck constant is   $\hbar=1$,
$|p\rangle=|n_{N-1}n_{N-2}\dots n_i\dots n_1n_0\rangle$,
$n_i=0,1$, and
\begin{equation}
\label{energy} E_p=\langle p|\mathbf{H}^0|p\rangle=
-\sum_{i=0}^{N-1}B^{z,0}_i(p)s^z_i(p).
\end{equation}
The dimensionless time $t$ in Eq.~(\ref{psi}) is
expressed in terms of the dimensional time $\bar t$ as
\begin{equation}
\label{time}
t={2e^2\over \hbar {\rm C_0}}\bar t.
\end{equation}
For ${\rm C}_0=500$ aF, the dimensionless time $t=1$
corresponds to the dimensional time $\bar t\approx 1$ ps, and one
dimensionless energy unit corresponds to 0.64 meV.

In the stationary field $B^{z,0}_i(p)$ and when $B^x_i=0$, the
coefficients $A_p(t)$ do not change when time $t$ changes from
$t'$ to $t''$, {\it {i.e.}} $A_p(t'')=A_p(t')$, while the
coefficients $c_p(t)$ evolve as
$c_p(t'')=c_p(t')e^{-iE_p(t''-t')}$. The representation of the
wave function in which the coefficients $A_p(t)$ are used is called
the ``interaction representation''. Since the coefficients
$A_p(t)$ are not changed during the free evolution of the system,
the interaction representation allows one to exclude from
consideration the dynamics associated with the evolution of the
phase of the wave function when no pulses are applied.

\section{One-qubit flip}
\label{sec:flip}
Let us discuss how to implement a resonant one-qubit rotation of
the $l$th qubit for the state $|p\rangle$ in our computer.
Initially at the time $t_0$
the $z$-component of the magnetic field is $B_l^{z,0}(p)$, and the
$B^x_l$ field is switched off, $B^x_l(t_0)=0$. The flip is implemented
in three steps.

(a) One changes the effective voltage $M_l(t)$ (below called
voltage) applied to the $l$th qubit, so that at the end of the
pulse of duration $\tau_1=t_1-t_0$ the magnitude of $M_l(t)$
becomes $M_l(t_1)={\cal M}_l$. If the transition is resonant
${\cal M}_l$ is defined by the resonant condition. [See
Eq.~(\ref{resonantM}) below.] From Eq.~(\ref{H01}) one can see
that after this pulse the $B^z$ field in the location of the $l$th
qubit is
\begin{equation}
\label{totalBz}
B^z_l(p,t_1)=B^{z,0}_l(p)-2{\cal M}_lU_{ll}.
\end{equation}
Since during the time of application of the first pulse the
$B^x_l$ field is turned off, the form of the first pulse is not
important, and the relevant parameter is the area
$\int^{t_1}_{t_0}M_l(t)dt$ of this pulse which determines phase
acquired by the wave function during this pulse. (b) One flips the
$l$th qubit by a rectangular $\pi$-pulse with the amplitude
$B^1_l$ and time duration $\tau_2=t_2-t_1=\pi/|B^1_l|$. The
$B_l^z$ field is not changed $B_l^z(p,t_2)=B_l^z(p,t_1)$ and
$M_l(t_2)=M_l(t_1)={\cal M}_l$. During this pulse the spin flips
through the angle $\pi$. (c) One restores the voltage applied to
the $l$th qubit from $M_l(t_2)={\cal M}_l$ to its original value
$M_l(t_3)=0$ during the time $\tau_3=t_3-t_2$.

First we will discuss the dynamics of the quantum computer during
implementation of the steps (a)-(c) in terms of the coefficients
$c_p(t)$, then we will formulate the result in the interaction
representation in terms of the coefficients $A_p(t)$. Let the
$l$th spin of a state $|p\rangle$ at the initial time $t_0$ be in
the state 0 and the state $|q\rangle$ be related to the state
$|p\rangle$ by a flip of the $l$th spin.

After the first pulse one has
\begin{equation}
\label{pulse1}
c_p(t_1)=e^{-iE_p\tau_1-i\theta^1_l(p)-i\varphi^1_l}c_p(t_0),~~~
c_q(t_1)=e^{-iE_q\tau_1-i\theta^1_l(p)+i\varphi^1_l}c_q(t_0).
\end{equation}
The total phase (for the states
$|p\rangle$ and $|q\rangle$) $\theta^1_l(p)$
and the phase $\varphi^1_l$ are defined by the time-dependent
component of the $B^z_l$ field in Eq.~(\ref{H01}) as
\begin{equation}
\label{varphi1}
\theta^1_l(p)=2\left[\int_{t_0}^{t_1}M_l(t)dt\right]
\sum_{i=0\atop i\ne\l}^{N-1}U_{li}s^z_i(p),~~~
\varphi^1_l=\left[\int_{t_0}^{t_1}M_l(t)dt\right]U_{ll}.
\end{equation}
At the end of the first pulse $M_l(t_1)={\cal M}_l$.

After the third pulse, the probability amplitudes become
\begin{equation}
\label{pulse3}
c_p(t_3)=e^{-iE_p\tau_3-i\theta^3_l(p)-i\varphi^3_l}c_p(t_2),~~~
c_q(t_3)=e^{-iE_q\tau_3-i\theta^3_l(p)+i\varphi^3_l}c_q(t_2).
\end{equation}
The phases are
\begin{equation}
\label{varphi3}
\theta^3_l(p)=2\left[\int_{t_2}^{t_3}M_l(t)dt\right]
\sum_{i=0\atop i\ne\l}^{N-1}U_{li}s^z_i(p),~~~
\varphi^3_l=\left[\int_{t_2}^{t_3}M_l(t)dt\right]U_{ll}.
\end{equation}
At the beginning of this pulse, one has $M_l(t_2)={\cal M}_l$, and
at the end of the pulse $M_l(t_3)=0$.

The action of the second rectangular pulse is described
by the following Schr\"odinger equation:
$$
i\dot c_p(t)=\left[E_p+b_l(p)+{\cal M}_lU_{ll}\right]c_p(t)
-{B^1_l\over 2}c_q(t)
$$
\begin{equation}
\label{eqMotion}
i\dot c_q(t)=\left[E_q+b_l(p)-{\cal M}_lU_{ll}\right]c_q(t)
-{B^1_l\over 2}c_p(t),
\end{equation}
where
\begin{equation}
\label{bl}
b_l(p)=2{\cal M}_l\sum_{i=0\atop i\ne\l}^{N-1}U_{li}s^z_i(p).
\end{equation}

In order to make the transition $|p\rangle\rightarrow |q\rangle$
resonant, the diagonal elements in Eq.~(\ref{eqMotion}) must be
equal to each other. The condition
\begin{equation}
\label{resonantE}
E_q-E_p=2{\cal M}_l^rU_{ll}
\end{equation}
defines the resonant value of the applied voltage ${\cal M}_l^r$.
We now calculate the energy difference $E_q-E_p$ using
Eq.~(\ref{energy}). Since the $B^{z,0}_l(p)=B^{z,0}_l(q)$ and
$s^{z}_l(p)=-s^{z}_l(q)$, the contribution to the energy
difference directly related to the flip of the $l$th qubit is
equal to $B^{z,0}_l(p)$. The states of other qubits with $i\ne l$
are the same for both states, $s^z_i(p)=s^z_i(q)$, but the fields
$B^{z,0}_i$, given by Eq.~(\ref{Bzi0p}), are different due to the
influence of the $l$th qubit measured by the matrix elements
$U_{li}$,
\begin{equation}
\label{BiDifference}
B^{z,0}_i(q)-B^{z,0}_i(p)=-\sum_{j=0\atop j\ne i}^{N-1}U_{ij}
[s^z_j(q)-s^z_j(p)]=U_{il}.
\end{equation}
Finally, we obtain
\begin{equation}
\label{eDifference}
E_q-E_p=B^{z,0}_l(p)-\sum_{i=0\atop i\ne\l}^{N-1}U_{li}s^z_i(p).
\end{equation}
From Eqs.~(\ref{resonantE}) and (\ref{eDifference}) the resonant
value of the applied voltage is
\begin{equation}
\label{resonantM}
{\cal M}_l^r={1\over 2U_{ll}}\left[
B^{z,0}_l(p)-\sum_{i=0\atop i\ne\l}^{N-1}U_{li}s^z_i(p)\right].
\end{equation}
Note, that the resonant value of the $B^z_l$ field
is not zero because of the constant inter-qubit interaction.
From Eqs.~(\ref{totalBz}) and (\ref{resonantM}) it is equal to
\begin{equation}
\label{resonantB}
B^z_l(p,t_1)=B^{z,0}_l(p)-2{\cal M}_l^rU_{ll}={1\over 2U_{ll}}
\sum_{i=0\atop i\ne\l}^{N-1}U_{li}s^z_i(p).
\end{equation}

To find the solution generated by the second pulse, introduce the
new coefficients $D_p(\tau)$ and $D_q(\tau)$ as
$$
c_p(t)=D_p(\tau)
e^{-i\left[E_p+b_l(p)+{\cal M}_lU_{ll}\right]\tau},
$$
\begin{equation}
\label{CD}
c_q(t)=D_q(\tau)
e^{-i\left[E_q+b_l(p)-{\cal M}_lU_{ll}\right]\tau},
\end{equation}
where $\tau=t-t_1$. Then, for the coefficients $D_p(\tau)$ and
$D_q(\tau)$ one obtains equations
$$
i\dot D_p(\tau)=-{B^1_l\over 2} D_q(\tau)e^{-i\Delta_l\tau},
$$
\begin{equation}
\label{eqMotionA}
i\dot D_q(\tau_2)=-{B^1_l\over 2}
D_p(\tau)e^{i\Delta_l\tau},
\end{equation}
where
\begin{equation}
\label{Deltaqp}
\Delta_l=E_q-E_p-2{\cal M}_lU_{ll}=
B^{z,0}_l(p)-\sum_{i=0\atop i\ne\l}^{N-1}U_{li}s^z_i(p)-
2{\cal M}_lU_{ll}.
\end{equation}
The condition
$\Delta_l=0$ is satisfied if the pulse is resonant
[see Eq.~(\ref{resonantM})].

If initially the $l$th qubit is in the state 0,
$D_p(0)=c_p(t_1)$ and $D_q(0)=0$, the solution of
Eq.~(\ref{eqMotionA}) is
$$
D_p(\tau_2)=c_p(t_1) \left[\cos\left({\lambda_l\tau_2\over
2}\right)+
i{\Delta_l\over\lambda_l}\sin\left({\lambda_l\tau_2\over 2}\right)
\right]e^{-i{\Delta_l\over 2}\tau_2},
$$
\begin{equation}
\label{solutionD}
D_q(\tau_2)=c_p(t_1)
i{B^1_l\over\lambda_l}\sin\left({\lambda_l\tau_2\over 2}\right)
e^{i{\Delta_l\over 2}\tau_2},
\end{equation}
where $\lambda_l=\sqrt{(B^1_l)^2+(\Delta_l)^2}$. From
Eq.~(\ref{solutionD}) one can see that if $\Delta_l=0$ and
$\tau_2=\pi/|B^1_l|$ there is a complete transition between the
states $|p\rangle$ and $|q\rangle$. Below we call $\Delta_l$ for
state $|p\rangle$ the detuning for this state.

We now express these results in the interaction representation in
terms of the coefficients $A_p(t)$. Let the state $|p\rangle$ with
$l$th qubit in the state $|0_l\rangle$ be populated at time
$t=t_0$. From the second expression in Eq.~(\ref{psi}) and the first
expression in Eq.~(\ref{pulse1}) one obtains
\begin{equation}
\label{CAt1}
c_p(t_0)=A_p(t_0)e^{-iE_pt_0},~~~
c_p(t_1)=A_p(t_0)e^{-iE_pt_1-i\theta^1_l(p)-i\varphi^1_l}.
\end{equation}
Taking into consideration Eqs.~(\ref{CD}),
(\ref{Deltaqp}), and (\ref{solutionD}), one has
$$
c_p(t_2)=A_p(t_0) \left[\cos\left({\lambda_l\tau_2\over 2}\right)+
i{\Delta_l\over\lambda_l}\sin\left({\lambda_l\tau_2\over 2}\right)
\right]e^{i\left\{-E_pt_2-\theta^1_l(p)-\varphi^1_l-
\left[b_l(p)+{\cal M}_lU_{ll}+ {\Delta_l\over
2}\right]\tau_2\right\}},
$$
\begin{equation}
\label{CAt2}
c_q(t_2)=A_p(t_0)
i{B^1_l\over\lambda_l}\sin\left({\lambda_l\tau_2\over 2}\right)
e^{i\left\{-E_qt_2-\theta^1_l(p)-\varphi^1_l-
\left[b_l(p)+{\cal M}_lU_{ll}+{\Delta_l\over 2}\right]\tau_2
+(\Delta_l+2{\cal M}_lU_{ll})t_2\right\}}.
\end{equation}
Finally, using Eq.~(\ref{pulse3}) one obtains in the interaction
representation
$$
A_p(t_3)=A_p(t_0) \left[\cos\left({\lambda_l\tau_2\over 2}\right)+
i{\Delta_l\over\lambda_l}\sin\left({\lambda_l\tau_2\over 2}\right)
\right]e^{i[\Theta_l(p)-\theta^3_l(p)-\varphi^3_l]},
$$
\begin{equation}
\label{interaction}
A_q(t_3)=A_p(t_0)
i{B^1_l\over\lambda_l}\sin\left({\lambda_l\tau_2\over 2}\right)
e^{i[\Theta_l(p)-\theta^3_l(p)+
\varphi^3_l+(\Delta_l+2{\cal M}_lU_{ll})t_2]},
\end{equation}
where the common phase
\begin{equation}
\label{ThetaP}
\Theta_l(p)=-\theta^1_l(p)-\varphi^1_l(p)-
\left[b_l(p)+{\cal M}_lU_{ll}+{\Delta_l\over 2}\right]\tau_2
\end{equation}
depends on the parameters of the first two pulses and
on the initial state $|p\rangle$.

\subsection{Calculation of the detuning $\Delta_l$}

Assume that initially there are two states, $|r\rangle$ and
$|R\rangle$, in the quantum register and assume
$\Delta_l=0$ for the state $|R\rangle$ and
$\Delta_l\ne 0$ for the state $|r\rangle$:
the $B^x$ pulse is resonant for the state $|R\rangle$
and nonresonant for the state $|r\rangle$.
The value of $\Delta_l$ for the state $|r\rangle$
can be found from Eqs.~(\ref{resonantM}) and (\ref{Deltaqp}),
\begin{equation}
\label{Delta1}
\Delta_l(r,R)=
-2\sum_{i=0\atop i\ne l}^{N-1}U_{li}[s^z_i(r)-s^z_i(R)].
\end{equation}
This equation can serve as a general definition of the detuning
$\Delta_l$ for an arbitrary state $|r\rangle$ expressed through
the eigenvalues of ${\mathbf S}^z_i$ for this state and the
eigenvalues of this operators for the state $|R\rangle$, for which
the transition is resonant. From Eq.~(\ref{Delta1}) one can see
that in general $\Delta_l(r,R)$ is independent of the static
voltages $m^0_j$, and for a resonant transition $\Delta_l(R,R)=0$.

\section{Protocol for creation of entangled state}

Let the initial state be the state $|0_{N-1}0_{N-2}\dots
0_10_0\rangle$. If $\eta\ll 1$ this is the ground state when the
applied $B^z$ field is oriented in the positive $z$-direction,
{\it i.e.} when $0\le m^0_i<1/2$. (See Fig.~2 for the case
$m^0_i=0$.) Using the Hadamard transform G$_{\rm H}$ we split the
ground state into two states
\begin{equation}
\label{VH} {\rm G}_{\rm H}|0_{N-1}0_{N-2}\dots 0_10_0\rangle=
{1\over\sqrt 2}(|0_{N-1}0_{N-2}\dots 0_10_0\rangle
+|0_{N-1}0_{N-2}\dots 0_11_0\rangle).
\end{equation}
Here and below we omit the total phase factor.
During the next step, we flip the first qubit in the
excited state and do not flip the same qubit in the
ground state to obtain the state
\begin{equation}
\label{2ndStep}
{1\over\sqrt 2}(|0_{N-1}0_{N-2}\dots 0_10_0\rangle
+|0_{N-1}0_{N-2}\dots 0_21_11_0\rangle).
\end{equation}
Repeating the latter procedure for the remaining $N-2$ qubits we
will obtain the entangled state:
\begin{equation}
\label{finalState}
{1\over\sqrt 2}(|0_{N-1}0_{N-2}\dots 0_10_0\rangle
+|1_{N-1}1_{N-2}\dots 1_11_0\rangle).
\end{equation}
We now define the parameters of the pulses required to implement
this protocol.

\subsection{Hadamard gate}

Assume that initially, at time $t_0$, there is only the ground
state $|0\rangle=|00\dots 00\rangle$ in the register. The Hadamard
transform is implemented using the sequence of the pulses (a)-(c)
described in Sec.\ref{sec:flip}. In step (b) instead of $\pi$
pulse we apply a $\pi/2$-pulse.

The form and duration of the first pulse are not important, since
they affect only the total phase of the wave function. What is
important is the value of the $B^z_0(0,t_1)$ field after this
pulse. To make this field equal to the resonant value (resonant
pulse) one should apply the voltage
\begin{equation}
\label{voltage0}
{\cal M}_0^r={1\over 2U_{00}}\left[
B^{z,0}_0(0)-\sum_{i=1}^{N-1}U_{0i}s^z_i(0)\right],
\end{equation}
where $s^z_i(0)=1/2$, and we used Eq.~(\ref{resonantM}). During
the second pulse one keeps the gate voltage constant. For the case
$m_j^0=0$ one has ${\cal M}_0^r=1/2$.
The second rectangular $\pi/2$-pulse with the amplitude $B^1_l$
has the duration $\tau_2=\pi/(2|B^1_l|)$. A specific value of
$B^1_l$ is not important. The only condition is $B^1_l\ne 0$.

During the third pulse the voltage is changed from
${\cal M}_0^r$ in Eq. (\ref{voltage0}) to its original value
which is equal to zero. To equalize the phases of two states
of the superposition, the condition
\begin{equation}
\label{phasesH}
-\varphi^3_0=
\varphi^3_0+[\Delta_0(0,0)+2{\cal M}_0^rU_{00}]t_2+{\pi\over 2}
\end{equation}
must be satisfied [see Eq. (\ref{interaction})]. Here
$\Delta_0(0,0)=0$. If we set initially $t_0=0$, then
\begin{equation}
\label{t2h}
t_2=\tau_1+\tau_2=\tau_1+{\pi\over 2|B^1_0|}.
\end{equation}
This gives us the last parameter of the Hadamard gate
\begin{equation}
\label{phasesH1}
\varphi^3_0=
-{\cal M}_0^rU_{00}\left(\tau_1+{\pi\over 2|B^1_0|}\right)
-{\pi\over 4}.
\end{equation}

\subsection{Conditional flip}

All other steps of the protocol implement flips
of the $1$st, $2$nd $\dots$, $(N-2)$th, $(N-1)$th qubits
in the excited state and suppress flips of these qubits
in the ground state. We will derive the parameters
required to implement this operation for the first qubit.
The parameters required to flip the other $N-2$ qubits can
be obtained in a similar way. Note that the parameters required
to flip different qubits are different even for a homogeneous
spin chain because the static $B^z$
field $B^{z,0}_i(R)$ is different for different qubits $i$ and
states $|R\rangle$, and because the phase factor
$\exp[i2{\cal M}_lU_{ll}t_2]$
in Eq.~(\ref{interaction}) depends on the history of the
excited state.

Before the pulses (a)-(c) are applied, there are two states in the
register $|r\rangle=|00\dots 00\rangle$ and $|R\rangle=|00\dots
01\rangle$. As for the Hadamard transform, the form and duration
of the first pulse is unimportant. The value of the voltage after
this pulse is
\begin{equation}
\label{voltage1} {\cal M}_1^r={1\over 2U_{11}}\left[
B^{z,0}_1(R)-\sum_{i=0\atop i\ne 1}^{N-1}U_{1i}s^z_i(R)\right].
\end{equation}
If $m^0_j=0$, then
\begin{equation}
\label{M01}
{\cal M}_1^r={1\over 2}+{U_{10}\over U_{11}}.
\end{equation}

The second rectangular $\pi$-pulse with the amplitude $B^1_1$ has
the duration $\tau_2=\pi/(|B^1_1|)$. In order to suppress the
nonresonant transition $|00\dots 000\rangle\rightarrow|00\dots
010\rangle$ from the ground state, the value of $|B^1_1|$ must
satisfy the $2\pi K$-condition~\cite{1997}
\begin{equation}
\label{2piK}
B^1_1=\pm {\Delta_1(r,R)\over\sqrt{4K^2-1}},~~~~K=1,2,\dots,
\end{equation}
where $B^1_1$ can be positive or negative.
For this particular pulse, using Eq.~(\ref{Delta1}), one obtains
$\Delta_1(r,R)=-2U_{10}$. If $B^1_1$ satisfies Eq. (\ref{2piK})
the value of the sine in Eq. (\ref{interaction}) becomes zero, and
the nonresonant transition is suppressed.

The voltage $M_1(t)$ is switched off, ${\cal M}_1^r\rightarrow 0$,
and the phase is corrected by the third pulse, for which the
condition
\begin{equation}
\label{phase20}
\Theta_1(r)-\varphi^3_1\left[1+{2\over U_{11}}
\sum_{i=0\atop i\ne 1}^{N-1}U_{1i}s^z_i(r)\right]+K\pi=
\varphi^3_1\left[1-{2\over U_{11}}
\sum_{i=0\atop i\ne 1}^{N-1}U_{1i}s^z_i(R)\right]
+\Theta_1(R)+2{\cal M}_1U_{11}t_2+{\pi \over 2}
\end{equation}
must be satisfied.
Here $t_2=T_1+\tau_1+\tau_2$, and $T_1$ is the time of the beginning
of the current three-pulse sequence.
Finally, the phase is
\begin{equation}
\label{phase30}
\varphi^3_1=
{\Theta_1(r)-\Theta_1(R)-2{\cal M}_1U_{11}t_2+(K-1){\pi\over 2}\over
2-{2\over U_{11}}
\sum_{i=0\atop i\ne 1}^{N-1}U_{1i}[s^z_i(R)-s^z_i(r)]}=
{\Theta_1(r)-\Theta_1(R)-2{\cal M}_1U_{11}t_2+(K-1){\pi\over 2}\over
2+{2U_{10}\over U_{11}}}.
\end{equation}

\section{Comments on practical implementation}

The protocol discussed in this paper provides the parameters for
exact implementation of entanglement in a chain of many coupled
SQUIDs. We demonstrated that the phase of the wave function can be
controlled by controlling the gate voltages. In a practical
situation one of the objectives can be the creation of the
entangled state with a relatively large constant of interaction
between qubits (but not using this state for implementation of
more complicated quantum algorithms). If the creation of the
entangled state is the purpose of an experiment, then the phases
of the entangled states can not be important. In this situation,
there is no need to control the phase $\varphi^3_l$, and the
important parameters are: (i) the value of the gate voltage ${\cal
M}_l$ applied to the qubit to be flipped, (ii) the amplitude
$B^1_l$ and (iii) the time-duration $\tau_2$ of $B^x$ pulse
[controlled by the magnetic flux $\Phi_l(t)$].

We now make some remarks concerning the specific field
configuration that we chose in this paper for implementation of
quantum logic. First, all qubits are initially placed in a
permanent static $B^z$ field, $B^{z,0}_j\ne 0$ created by static
voltages $m^0_j$ and the inter-qubit interaction. Before a flip of
the $l$th qubit by a $B^x$ pulse, the voltage applied to this
qubit is switched to the resonant value, and after the $B^x$-pulse
this voltage is switched to its initial value. As follows from the
results of this paper, the nonzero static $B^{z,0}_j$ field
introduces an additional parameter, or degree of freedom: by
changing this field it is possible to control the phase of the
wave function.

The qubit rotations are implemented by the rectangular
$B^x_l$-pulses because during implementation of these pulses the
effective $B^z_l$ field in the location of the $l$th qubit is not
equal to zero for some quantum states of the superposition. That
is why the result of the action of a $B^x$-pulse on the states
with nonzero $B^z_l$ field depends on the form of this pulse. The
analytic solution is known for a rectangular pulse, so the
rectangular form of the $B^x_l$-pulse is the simplest possible
choice that allows one to analyze the basic properties of the
system and optimize the parameters of the pulses.

The interaction representation used in this paper
for modeling of quantum logic is employed because it
effectively eliminates from consideration free quantum
dynamics of the quantum states of the superposition.
Namely, since different states of the superposition have different
energies $E_p$, they acquire different phase factors
even in the case when the applied fields are not changed.
The interaction representation allows one to choose the
static voltages $m^{0}_i$ arbitrarily. In particular, one can
choose $m^0_i=0$. (See, for example, Fig. 2 of this paper.)
If $n^0_i=0$, this would allow one to decrease the decoherence
effect of the gate because the gate voltage is switched
on and off only during the logic operation on a specific
qubit. Also the time between the pulses can be
chosen arbitrarily because the wave amplitudes
in the interaction representation are not changed during
this time.

\section{Long-range interaction}

\label{sec:numerical} For two states in the register, the system
with long-range interaction between  qubits is exactly solvable.
However, the universal quantum computation involving an arbitrary
possible number of states in the quantum register is supposed to
be implemented by taking into consideration only nearest neighbor
interaction. The long-range interaction in the general case
results in generation of error states. These errors are small only
when the value of $\eta$, characterizing the coupling strength, is
small. For example, the value $\eta=0.1$ can be considered as
large because the long-range interaction produces the error of the
order $\eta=0.1$ in the probability amplitude and phase of the
wave function. In order to show this, consider
Eq.~(\ref{interaction}). Due to Eq.~(\ref{Delta1}) the
contribution of $(l-2)$th and $(l+2)$th qubits to the detuning
$\Delta$ is of order $\eta^2$; the contribution of $(l-3)$th and
$(l+3)$th qubits to the detuning $\Delta$ is of order $\eta^3$,
and so on. If the parameters of the pulses are calculated by
taking into consideration only nearest neighbors, instead of the
resonant transition with $\Delta=0$ one has an almost resonant
transition (for $\eta\ll 1$) with the nonzero detuning
$\Delta\sim\eta^2$ created by $(l-2)$th and $(l+2)$th qubits.
Since in Eq.~(\ref{interaction}) $\lambda_l\sim\eta$, the relation
$\Delta_l/\lambda_l$ is of order of $\eta$; the value of the sine
(for almost resonant transition) is of order of unity; so that the
error in the probability amplitude and phase of the wave function
is of order $\eta$.

\begin{figure}
\centerline{\includegraphics[width=9cm,height=9cm]{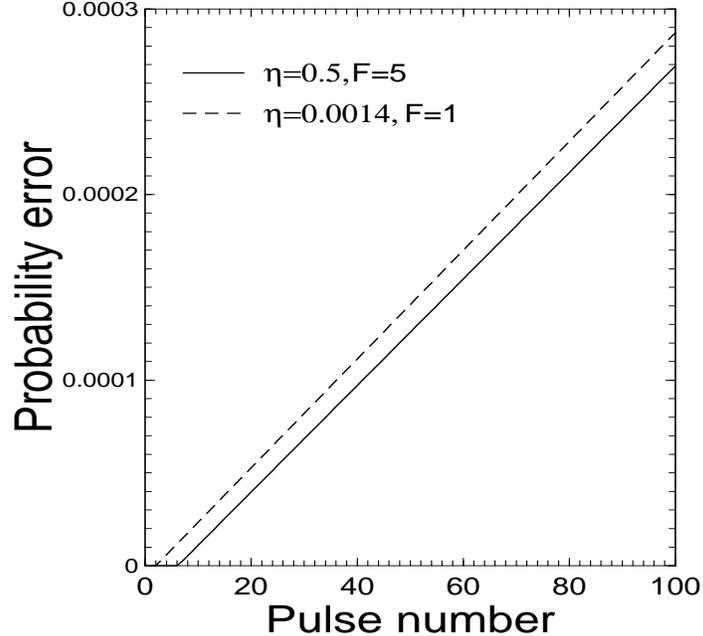}}
\vspace{-5mm} \caption{The probability error as a function of
$B^x$ pulse number. $N=100$, $K=1$; $m^0_j=0$, $j=0,1,\dots,N-1$.}
\label{fig:3}
\end{figure}

In order to calculate the error introduced by the long-range
interaction, we numerically simulated the creation of entanglement.
The parameters of the pulses were calculated by taking
into consideration only $F$ nearest neighbors, $F=1,2,\dots$.
We define the probability error as
\begin{equation}
\label{probabilityError} {\cal P}(t)=\left||A_r(t)|^2-{1\over
2}\right| + \left||A_R(t)|^2-{1\over 2}\right|,
\end{equation}
and plot it as a function of the number of $B^x$ pulses in Fig.~3.
The total number of $B^x$
pulses is equal to the number of qubits $N$ in the
chain. In Eq.~(\ref{probabilityError}) $|r\rangle$ is the ground state
and $|R\rangle$ is the excited state.

The probability error ${\cal P}(t)$ in Fig.~3 generated by one
pulse is equal, on average, to $2.87\times 10^{-6}$ ($9.77\times
10^{-4}$) for $F=5$ and $\eta=0.5$; and ${\cal P}=2.93\times
10^{-6}$ ($1.96\times 10^{-6}$) for $F=1$ and $\eta=0.0014$. In
brackets we indicate the values calculated using the analytical
estimates presented in the beginning of this section. Namely, the
error in the probability amplitude is of the order $\eta^F$, and
the error in the probability is of order $\eta^{2F}$. The poor
correspondence between estimated and numerical results for
$\eta=0.5$ follows from the fact that for relatively large $\eta$
the off-diagonal matrix elements in the matrix $U_{ij}$ decrease
faster than $\eta^{|i-j|}$, $i\ne j$.

As follows from Fig.~3, in order to decrease the error one can (i)
take into consideration a larger number of nearest neighbors or
(ii) decrease the coupling constant $\eta$. The results presented
in Fig.~3 can be used to estimate the probability error in more
complex protocols with the same values of $F$ and $\eta$, since
one can assume that this error mostly depends on the number of
$B^x$ pulses and not on a particular protocol.

\begin{figure}
\centerline{\includegraphics[width=9cm,height=9cm]{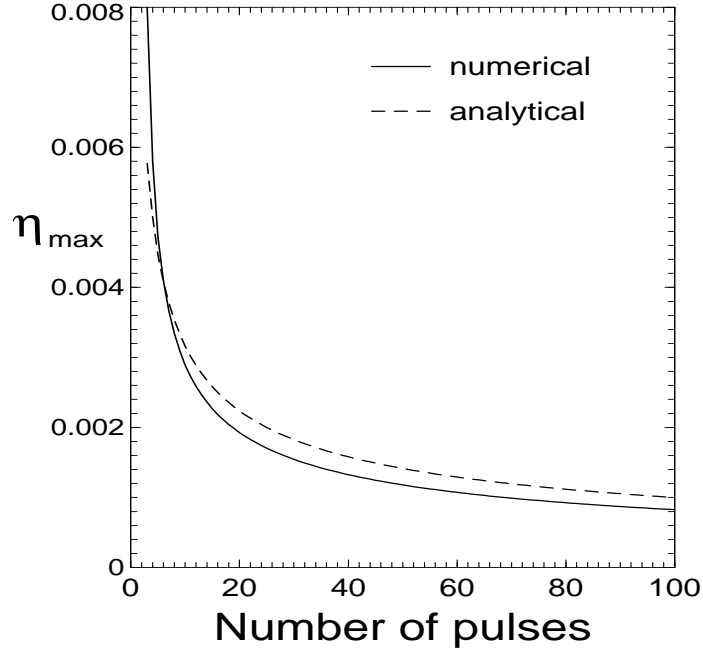}}
\vspace{-5mm}
\caption{The maximum value of $\eta$ ($y$-axis) required to
implement the entanglement protocol using $N$ pulses ($x$-axis),
with the probability error
less than ${\cal P}_0=10^{-4}$. $F=1$,
$K=1$; $m^0_j=0$, $j=0,1,\dots,N-1$. The solid curve
is obtained using Eq.~(\ref{interaction}) and
the dashed curve is obtained using Eq.(\ref{epsilonP}).}
\label{fig:4}
\end{figure}

As shown in this paper, the effects of the long-range interaction
can be completely compensated by choosing the optimal values of
the $B^z$ and $B^x$ fields when the number of useful states in the
quantum register is equal to 2. In the general case of an arbitrary
(possible)
number of states in the register, only the nearest neighbors can
be taken into consideration, and the long-range interaction can be
suppressed only by decreasing the value of $\eta$. In Fig.~4 we
plot the maximum value $\eta_{\rm max}$ required to implement the
protocol for creation of entanglement using $N$ pulses with the
probability error less than ${\cal P}_0=10^{-4}$ when only the
nearest neighbors are taken into consideration, $F=1$.

The value of $\eta_{\rm max}$ in Fig.~4 can be estimated analytically.
As follows from Fig.~3 the error accumulates linearly
with the number of pulses, so that we can write
\begin{equation}
\label{PEpsilon}
{\cal P}_0=N{\cal P}^\prime=N\eta_{\rm max}^2,
\end{equation}
where ${\cal P}^\prime$ is the error generated by one pulse.
From Eq.~(\ref{PEpsilon}) the maximum
value of $\eta$ required to implement an $N$-pulse protocol
with the accuracy ${\cal P}_0=10^{-4}$ is
\begin{equation}
\label{epsilonP}
\eta_{\rm max}=\sqrt{{{\cal P}_0\over N}}={0.01\over\sqrt N}.
\end{equation}
From Fig.~4 one can see that our analytical estimate for
$\eta_{\rm max}$ is close to the exact result obtained using
numerical simulations.

Decreasing $\eta$ decreases the error.
On the other hand, since $\eta$ defines the clock
speed of the quantum computer (the time of implementation
of one $B^x$ pulse is of order of $\pi/\eta$), decreasing
$\eta$ slows the computer down. This can lead to an
accumulation of errors introduced by the environment.
The optimal value of $\eta$ can be estimated by taking
into consideration both the effect of the long-range
interaction and the influence of the environment.

\section{Conclusion}

We considered in this paper an optimal implementation of quantum
logic operations for a scalable superconducting quantum computer
with constant inter-qubit interaction. The protocol for creation
of entanglement with arbitrary number of qubits was analyzed in
detail, for a relatively large interaction constant. A possible
application of our approach for implementing universal quantum
logic for more complex algorithms by decreasing the coupling
constant and, correspondingly, decreasing the clock speed was
discussed. The errors introduced by the long-range interaction for
the universal logic gates are estimated analytically and
calculated numerically. The coherent charge oscillations in this
model have already been observed experimentally~\cite{pashkin03}.
A further feasible accomplishment would be the experimental
creation of entanglement in a system with three qubits. A
demonstration of the entanglement in the potentially scalable QC
architecture with the superconducting qubits would be an important
step toward the experimental implementation of a scalable QC with
many qubits.

\section*{Acknowledgments}

We thank G.D. Doolen for discussions. This work  was supported by
the Department of Energy under the contract W-7405-ENG-36 and DOE
Office of Basic Energy Sciences, by the National Security
Agency (NSA) and Advanced Research and Development Activity (ARDA)
under Army Research Office (ARO) contract \# 707003.

\appendix
\section{Creation of entanglement with two qubits}
Let us calculate the parameters required to create the
entanglement in the chain of two qubits starting from the state
$|00\rangle$ and using the parameters from the experimental 
system~\cite{pashkin03}. We will not calculate the phase
$\varphi^3_l$ required to implement the phase correction, since it
depends on particular forms of the gate pulses $M_l(t)$, and
calculate the parameters required to create the entanglement
without phase correction. There are three devices attached to
$i$th ($i=0,1$) SQUID: superconducting electrode with the
capacitance ${\rm C}_{Ji}$, electrostatic gate with the
capacitance ${\rm C}_{gi}$, and a probe with the capacitance ${\rm
C}_{bi}$. We assume that during implementation of the protocol the
probe voltages $V_{bi}$ are switched off, so that the charges
$Q_{bi}={\rm C}_{bi}V_{bi}$ induced by the probes are equal to
zero. The capacitances are
\begin{equation}
\label{capacitances0}
{\rm C}_{J0}=620~{\rm aF},~~~ {\rm C}_{g0}=0.60~{\rm aF},~~~
{\rm C}_{b0}=41~{\rm aF},~~~
{\rm C}_0={\rm C}_{J0}+{\rm C}_{g0}+{\rm C}_{b0}=661.6~{\rm aF}.
\end{equation}
\begin{equation}
\label{capacitances1}
C_{J1}=460~{\rm aF},\qquad C_{g1}=0.61~{\rm aF},\qquad
C_{b1}=50~{\rm aF}, \qquad {\rm C}_1=510.61~{\rm aF}.
\end{equation}
The coupling capacitance ${\rm C}_c$ and the coupling
constant $\eta$ are
\begin{equation}
\label{coupling0}
{\rm C}_c=34~{\rm aF},\qquad
\eta={{\rm C}_c\over {\rm C}_0}=
{34\over 661.6}\approx 5.139\times 10^{-2}.
\end{equation}
The coupling matrix $U_{ij}$, $i,j=0,1$, is given by
Eq.~(\ref{J2}), where
\begin{equation}
\label{couplingA}
a={{\rm C}_1\over {\rm C}_0}=
{510.61\over 661.6}\approx 0.7718.
\end{equation}
The dimensionless parameters $\eta$ and $a$
and the dimensionless static voltages $m^0_0$ and $m_0^1$
completely define the parameters of the protocol.

The value of the voltage ${\cal M}_0$
applied to the zeroth qubit is given by Eq.~(\ref{resonantM}),
\begin{equation}
\label{M0}
{\cal M}^r_0={1\over 2}-m_{0}^0-{U_{01}\over U_{00}}m^0_{1}.
\end{equation}
The amplitude $B^1_0$ of the applied $B^x$ field is arbitrary, and
the time-duration of this pulse is
\begin{equation}
\label{tau0}
\tau_2={\pi+2\pi k\over 2|B^1_0|},\qquad k=0,1,2,\dots.
\end{equation}

The gate voltage ${\cal M}_1$ for the second pulse is
\begin{equation}
\label{M1}
{\cal M}^r_1={1\over 2}-m_{1}^0+{U_{10}\over U_{11}}(1-m^0_{0}).
\end{equation}
The magnitude of $\Delta_1$ calculated using Eq.~(\ref{Delta1}) is
$-2U_{01}$, and the amplitude of $B^x$ field is
\begin{equation}
\label{B1}
B^1_1={2U_{01}\over \sqrt{4K^2-1}},\qquad K=1,2,\dots.
\end{equation}
The time-duration of the second $B^x$ pulse
is
\begin{equation}
\label{tau1}
\tau_2={\pi+2\pi k\over |B^1_1|},\qquad k=0,1,2,\dots.
\end{equation}

\section{Creation of entanglement with three qubits}

The matrix $U_{ij}$ for the system of three identical qubits
is given by Eq.~(\ref{J3}). The protocol presented in this Appendix
is also valid for a system with nonidentical qubits.
The gate voltage ${\cal M}_0$ for the first $B^x$ pulse is
\begin{equation}
\label{M03}
{\cal M}^r_0={1\over 2}-m_{0}^0-{U_{01}\over U_{00}}m^0_{1}
-{U_{02}\over U_{00}}m^0_{2}.
\end{equation}
The magnitude of $B^1_0$ is arbitrary, and $\tau_2$ is given by
Eq.~(\ref{tau0}).

The gate voltage ${\cal M}_1$ for the second pulse is
\begin{equation}
\label{M13}
{\cal M}^r_1={1\over 2}-m_{1}^0+{U_{10}\over U_{11}}(1-m^0_{0})
-{U_{12}\over U_{11}}m^0_{2}.
\end{equation}
The magnitudes of $\Delta_1=-2U_{01}$,
$B^1_1$, and $\tau_2$ for the second
pulse are given by the same formulas (\ref{B1}) and (\ref{tau1})
as for the case of two qubits.

The gate voltage ${\cal M}_2$ for the third pulse is
\begin{equation}
\label{M23}
{\cal M}^r_2={1\over 2}-m_{2}^0+{U_{20}\over U_{22}}(1-m^0_{0})
+{U_{21}\over U_{22}}(1-m^0_{1}).
\end{equation}
The detuning $\Delta_2$ for the third pulse is $-2U_{20}-2U_{21}$,
and the amplitude of $B^x$ field is
\begin{equation}
\label{B23}
B^1_2={2U_{20}+2U_{21}\over \sqrt{4K^2-1}}, \qquad K=1,2,\dots,
\end{equation}
The time-duration of the third $B^x$ pulse is
\begin{equation}
\label{tau2}
\tau_2={\pi+2\pi k\over |B^1_2|},\qquad k=0,1,2,\dots.
\end{equation}
{}

\begin{thebibliography}{}

\bibitem{makhlin01}
Y. Makhlin, G. Sch\"on,
and A. Shnirman, Rev. Mod. Phys. {\bf 73},
357 (2001).

\bibitem{chiorescu03}
I. Chiorescu, Y. Nakamura,
C.J.P.M. Harmans, and J. E. Mooij,
Science {\bf 299}, 1869 (2003).

\bibitem{nakamura99}
Y. Nakamura,
Yu. A. Pashkin, and J. S. Tsai, Nature {\bf 398}, 786 (1999).

\bibitem{pashkin03}
Yu. A. Pashkin, T. Yamamoto, O. Astafiev, Y. Nakamura,
D. V. Averin, and J. S. Tsai, Nature {\bf 421}, 823 (2003).

\bibitem{ramos}  
R. C. Ramos, F. W. Strauch, P. R. Johnson,
A. J.  Berkley, H. Xu, M. A. Gubrud, J. R. Anderson,
C. J. Lobb, A. J. Dragt, and F. C. Wellstood, 
IEEE Trans. on Appl. Supercond. {\bf 13}, 994 (2003).

\bibitem{martinis02} J. M. Martinis,
S. Nam, J. Aumentado, and C. Urbina,
Phys. Rev. Lett. {\bf 89}, 117901 (2002).

\bibitem{yu02} Y. Yu, S. Han, X. Chu, S. Chu. and Z. Wang,
Science {\bf 296}, 889 (2002).

\bibitem{makhlin99}
Y. Makhlin,
G. Sch\"on, and A. Shnirman, Nature {\bf 386}, 305 (1999).

\bibitem{plastina01} F. Plastina, R. Fazio, and G. M. Palma,
Phys. Rev. B {\bf 64}, 113306 (2001).

\bibitem{you01} J. Q. You, C.-H. Lam, and H. Z.  Zheng,
Phys. Rev. B {\bf 63}, 180501 (2001).

\bibitem{you02} J. Q. You, J. S. Tsai, and F. Nori,
Phys. Rev. Lett. {\bf 89}, 197902 (2002).

\bibitem{blais03} A. Blais,
A. M. van den Brink, and A. M. Zagoskin,
Phys. Rev. Lett. {\bf 90}, 127901 (2003).

\bibitem{johnson03} P. R. Johnson, F. W. Strauch,
A. J. Dragt, R. C. Ramos, C. J. Lobb, J. R. Anderson,
and F. C. Wellstood, Phys. Rev. B {\bf 67}, 020509 (2003).

\bibitem{plastina03} F. Plastina and G. Falci,
Phys. Rev. B {\bf 67}, 224514 (2003).


\bibitem{pashkin1} Y. Nakamura, Yu. A. Pashkin, and J. S. Tsai,
Phys. Rev. Lett. {\bf 87}, 246601 (2001). 

\bibitem{pashkin2} Y. Nakamura, Yu. A. Pashkin, and J. S. Tsai
in: {\it Macroscopic Quantum Coherence and Quantum Computing},
Eds. D. V. Averin, B. Ruggiero, and P. Silvestrini, p. 17
(Kluwer Academic/Plenum Publishers, New York, 2001)

\bibitem{stacked} C. Granata, V. Corato, A. Monaco, B. Ruggiero, 
M. Russo, and P. Silvestrini,
Appl. Phys. Lett. {\bf 79}, 1145 (2001).

\bibitem{bruder93} C. Bruder, R. Fazio, and G. Sch\"on,
Phys. Rev. B {\bf 47}, 342 (1993).

\bibitem{glazman97} L. I. Glazman and A. I. Larkin,
Phys. Rev. Lett. {\bf 79}, 3736 (1997).
%
\bibitem{1997} G. P. Berman, D. K. Campbell,
and V. I. Tsifrinovich, Phys. Rev. B {\bf 55}, 5929 (1997).
%
\end{thebibliography}
\end{document}